\documentclass[reprint, showpacs, preprintnumbers, amsmath, amssymb, aps, nofootinbib]{revtex4}

\usepackage{upgreek}

\usepackage[T1]{fontenc}
\usepackage{natbib}
\usepackage[normalem]{ulem}
\usepackage{amssymb,amsmath,mathrsfs,amsbsy}
\usepackage{multirow,xcolor,soul,graphicx}

\newcommand{\BF}[1]{\mathbf{#1}}
\newcommand{\mr}[1]{\mathrm{#1}}

\def\bu{\mathbf{u}}
\def\hbu{\hat{\bu}}

\def\C{^\circ \mathrm{C}}

\begin{document}

\title[Leidenfrost Flows]{Leidenfrost Flows : instabilities and symmetry breakings}
%
\author{E. Yim$^{1,\dagger}$, A. Bouillant$^{2,3,\dagger}$, D. Qu{\'{e}}r{\'{e}}$^{2,3}$, F. Gallaire$^{1}$}
\affiliation{$^{1}$LFMI, \'Ecole Polytechnique F\'ed\'erale de Lausanne, CH-1015 Lausanne, Switzerland \\ 
$^{2}$LadHyX, \'Ecole polytechnique, 91128 Palaiseau, France \\
$^{3}$PMMH, PSL-ESPCI, CNRS-UMR 7636, 75005 Paris, France\\
$\dagger$ Authors contributed equally to this work }


\date{\textbf{Received:} XX 2021; \textbf{Revised:} XX XX 2021; \textbf{Accepted:} XX XX 2022}

\begin{abstract}
 Leidenfrost drops were recently found to host strong dynamics. In the present study, we investigate both experimentally and theoretically the {flows structures and stability} inside a Leidenfrost water drop as it evaporates, starting with a large puddle. As revealed by infrared mapping, the drop base is warmer than its apex by typically 10$^{\circ}$C, which is likely to trigger bulk thermobuoyant flows and Marangoni surface flows. Tracer particles unveil complex and strong flows that undergo successive symmetry breakings as the drop evaporates. We investigate the linear stability of the baseflows in a non-deformable, quasi-static, levitating drop induced by thermobuoyancy and effective thermocapillary surface stress, using only one adjustable parameter. The stability analysis of nominally axisymmetric thermoconvective flows, parametrized by the drop radius $R$, yields the most unstable, {thus, dominant, azimuthal modes (of wavenumber $m$).  Our theory predicts well the radii $R$ for the mode transitions and cascade with decreasing wavenumbers from $m=3$, $m=2$, down to $m=1$  (the eventual rolling mode that entails propulsion) as the drop shrinks in size}.  The effect of the escaping vapor is not taken into account here, which may further destabilize the inner flow and couple to the liquid/vapor interface to give rise to motion \citep{Bouillant2018a,brandao2020spontaneous}.\end{abstract}
\maketitle

\section{Introduction}
A water drop can levitate above a hot surface provided the solid temperature exceeds the boiling temperature of the liquid by typically 100$^{\circ}$C. This effect was first reported in 1756 by J.G. Leidenfrost \citep{Leidenfrost1756} and it has been ever since a source of scientific curiosity.
Evaporation produces a vapor film underneath the drop with typical thickness 50 $\mu$m that thermally insulates the liquid from its substrate and suppresses boiling. The drop can thus survive a few minutes to plate temperatures as high as 300$^{\circ}$C. Levitation also prevents the liquid from wetting the surface. As a consequence, a drop adopts a quasi-spherical shape when its radius $R$ is smaller than the capillary length $\ell_c = \sqrt{{\sigma_0}/{\rho_0 g}}$ $-$ where $\sigma_0$ is the surface tension, $\rho_0$ is the liquid density and $g$ is the acceleration of gravity, that is $\ell_c \approx 2.5$~mm for water at 100$^{\circ}$C $-$ while it gets flattened by gravity to a height 2$\ell_c$ when $R > \ell_c$. The absence of contact also produces a (quasi-)frictionless situation by which the liquid drop becomes highly mobile. 
Thermal and mechanical insulation have contrasted consequences.
On the one hand, above the Leidenfrost temperature, vapor compromises the liquid cooling properties and pilots the transition from nucleate to film boiling, unwelcome in nuclear reactor or in metallurgy (loss of control in quenching processes). On the other hand, it produces a purely non-wetting situation, which can serve as a canonical model for superhydrophobicity. Recent studies have illustrated the richness of spontaneous dynamics related to Leidenfrost drops \citep{Quere2013}, including oscillations \citep{Garmett1878,Holter1952,Brunet2011,Bouillant2020}, bouncing \citep{Celestini2012,Waitukaitis2017} and directed propulsion on asymmetrically textured \citep{Linke2006,Cousins2012} or on non-uniformly heated surfaces \citep{Sobac2017,Bouillant2021}, which could be exploited in the design of micro-reactors \citep{Raufaste2016}, of heat pipes and exchangers, as well as for lab-on-a-chip technologies.\\
Among those dynamics is the ability of Leidenfrost droplets to self-rotate and self-propel, even in the absence of external fields \citep{Bouillant2018a}. Despite the early discovery of the Leidenfrost effect, the existence of strong inner flows has only been reported recently, with velocities as high as a few cm/s and whose morphology switches from four counter-rotating swirling cells, preserving the axial symmetry, to an unique asymmetric swirl as evaporation proceeds \citep{Bouillant2018b}. The {eventual} solid-like asymmetric rotation {comes with} a tilt at the droplet base by an angle $\alpha$ of typically a few milliradians { \citep{Bouillant2018a}}, {producing accelerations that scale as} $a \sim \alpha g$ \citep{Dupeux2013} of a few tens of mm/s$^2$. {The internal rolling actually couples with the vapor cushion in a feedback loop to sustain the self-rolling and self-propelling motions \citep{brandao2020spontaneous}, which thereby explains the intrinsic mobility of Leidenfrost droplets}.
The origin of inner flows {as well as their structure remain} however unclear. {They may arise} from: {i) a thermal {scenario}: the base of the drop is close to the hot plate while its apex is exposed to cooler air. Temperature differences in the liquid may give birth to convective flows, either driven by thermocapillary effects, known as Marangoni effect (variations of the surface tension along the temperature gradient as in \cite{Scriven60}) or driven in the liquid bulk by thermobuoyant effects, known as Rayleigh-B\'enard convection (resulting from the thermal expansion of the liquid); ii) a hydrodynamic {scenario}: vapor produced at the drop base escapes through the subjacent film exerting a viscous radial stress that may draw liquid with it. Both scenarios \textit{a priori} preserve axisymmetry, which is contradicted by experimental observations. Indeed, Particle induced velocimetry (PIV) has revealed that internal flow structures evolve in time and therefore with the drop geometry \citep{Bouillant2018a}, suggesting a symmetry selection mechanism as the drop shrinks in size.  
Similar selection mechanism has been reported 
in a non-levitating drop deposited on a warm plate \citep{Tam09,Dash14}. In this close configuration, the vapor cushion is suppressed and the origin of the internal dynamics is purely thermal.  Temperature gradients being greater in Leidenfrost drop than for sessile drop on warm plates,  we anticipate enhanced thermoconvection,  prompting us to focus on the thermal scenario (i) rather than on the hydrodynamics scenario relying on the escaping vapor (ii).
We therefore address the stability and the symmetry of thermo-induced flows inside a Leidenfrost drop as it evaporates, neglecting the hydrodynamic effect from the escaping vapor and without explicitly modelling the surrounding gas.  We aim at predicting the radii at which are expected the successive symmetry breakings  starting with $R<4\ell_c$ (to prevent chimney formation), that is in the regime where drops are flatten by gravity ($R>\ell_c$), while approaching the regime where drops get quasi-spherical ($R<\ell_c$). 
Note that we do not specifically explore the transition to the rolling mode $m=1$ as in \cite{Bouillant2018a,brandao2020spontaneous} but we explore higher modes $m>1$ observed in large puddles $R>\ell_c$. 
In this regime,  vapors carve a blister underneath the liquid, whose amplitude increases with $R$ and reaches the entire height of the puddle, $2\ell_c$, when $R\lesssim 4\ell_c$.  Viscous entrainment being markedly weakened along this vapor pocket, vapors would only draw liquid along the very narrow, peripheral neck \citep{Pomeau2012,Burton2012,Sobac2014}. The peculiar geometry of the vapor cushion underneath large puddles undermines the hydrodynamic scenario, corroborating our assumptions.
Hydrodynamic effects of the vapor cushion may however shift our prediction for the bifurcation radii, especially when $R\rightarrow \ell_c$. As shown by \cite{brandao2020spontaneous} in the case of quasi-spherical drops, the internal rolling couples with the vapor cushion in a feedback loop, which sustains the self-rolling and self-propelling motions. Such a coupling could also reshape the vapor cushion when $R>\ell_c$, with consequences on the drop mobility that are not captured by our model.\\
The selection mechanism for the flow symmetry is a common feature in both buoyancy and Marangoni instabilities. These instabilities are known to be very sensitive to the geometry and confinement; the preferred discrete azimuthal mode wavenumber decreases with the liquid domain size.}
The stability analysis on the Marangoni-Rayleigh-B\'enard convection has been extensively studied for a rectangular and cylindrical domain with various boundary conditions \cite{Pearson58,Nield64,Vrentas81,Rosenblat82,Kuhlmann93,Johnson99}. Yet, the stability of a free liquid drop subjected to a vertical temperature gradient has heretofore not received the same attention. 
Therefore, Leidenfrost drops constitute a toy model where the evaporation-driven confinement enables to {quasi-steadily} sweep the states in a non-wetting drop heated from below. We restrict our parametric study in $R$ to the limit of $R < 5$~mm to prevent Leidenfrost chimneys and pulsating stars to appear \citep{Quere2013}.
Leaning on the experimental observations, we develop a numerical model, which implicitly decouples the evaporation timescale from the meanflow evolution timescale. We thus look for the stability of the nominally axisymmetric thermo-convective baseflow in Leidenfrost drops in order to explain the symmetry breaking from 4 to 1 convective cells reported in \cite{Bouillant2018b}, as well as the prior transition from 6 to 4 cells.\\
We first characterize the successive symmetry breakings in the internal flows (\S \ref{sec:Exp}) and extract from experiments physical quantities relevant to the problem {such as the temperature difference at the drop interface}.
The governing hydrodynamic and thermal equations, as well as the linear stability analysis are then presented in \S \ref{sec:Theory} and the results are shown in \S \ref{sec:result}. We obtain the baseflow generated by the stratification within the liquid and search for the effective Marangoni number, which best captures our experimental observations, such as the surface temperature and the velocity field.
Then, a stability analysis is carried out for different drop radii $R$. We compare for a given $R$ the stability properties of each symmetry breaking mode, the mode with the highest growth rate -- the most unstable mode -- being expected to dominate the flow structure. 
Our study predicts the successive inner flow symmetries as drops shrink in size. It also provides the critical radii for the modes transition in quantitative agreement with observations. We eventually discuss and compare the numerical outcomes to experiments, and add a few concluding remarks in \S \ref{sec:discuss}.

\section{Experimental observations}
\label{sec:Exp}
\subsection{Quasi-steady state assumption}
\label{sec:quasisteady}
\begin{figure}[!htb]
\centering
\includegraphics[width=0.7\textwidth]{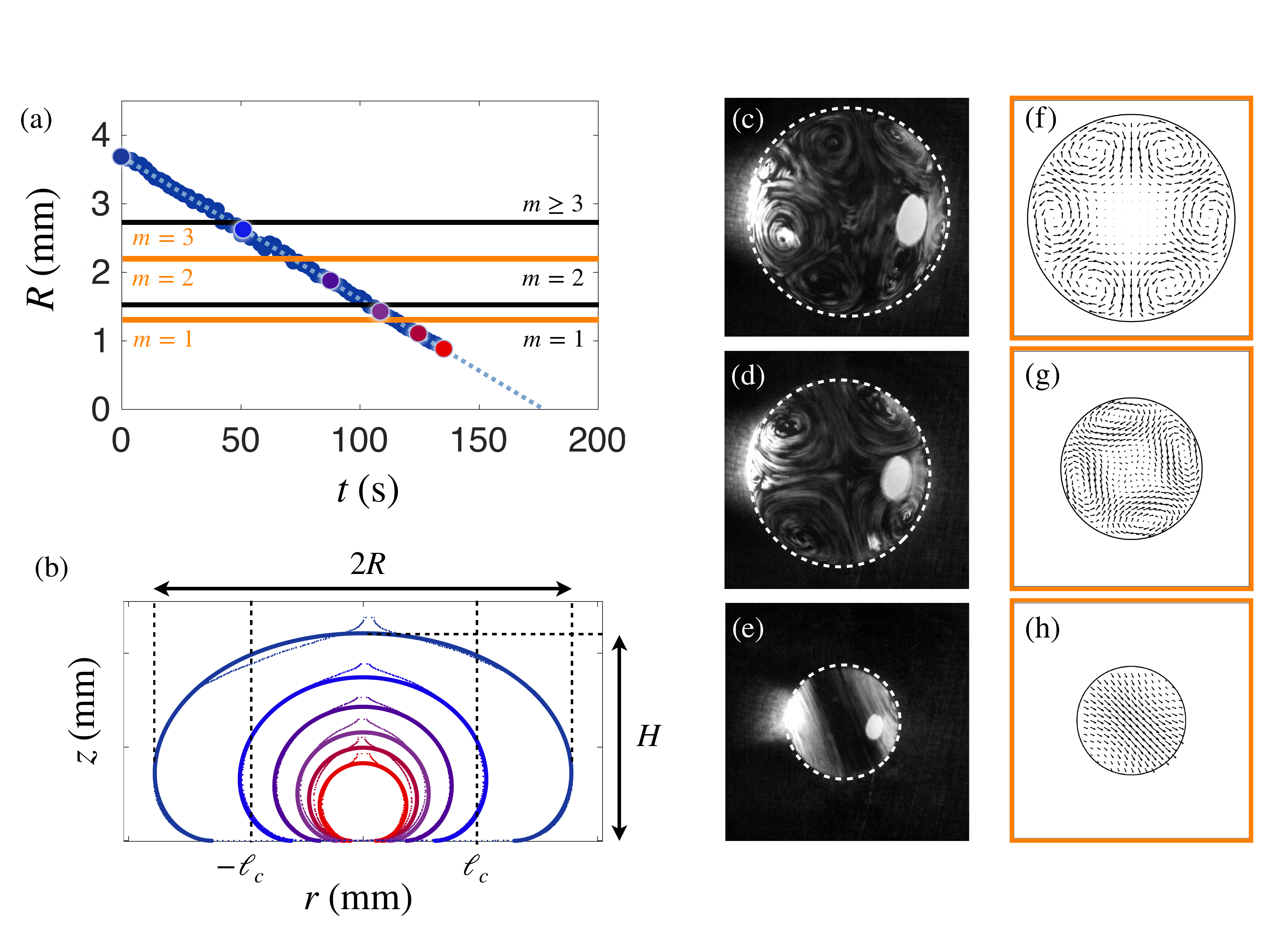}
\caption{\label{fig:rvst} (a) Radius $R$ of a Leidenfrost drop {levitating on a plate heated at} 350$^{\circ}$C as a function of time $t$. $R$ decreases as $R(t)=R_0(1-t/\tau_0)$ (eq.\ref{eq:fitRT}), plotted as dotted line, denoting $R_0 = 3.7\pm 0.1$~mm as the initial radius and $\tau_0 = 176.5$~s as the lifetime. 
(b) Drop shape for some radii $R\in[0.9 ;4.5]$~mm readable in (a) (see colored dots). Simulated shapes (full lines), obtained by numerically integrating (\ref{eq:shape}), are compared to experimental ones (dotted lines), obtained for side-viewed drops pinned with a needle. $\ell_c$ is the capillary length.
Surface flows, viewed from the top, successively self-organize into (c) 6 counter-rotating cells (mode $m=3$), (d) 4 counter-rotating cells (mode $m=2$), and eventually a unique rolling cell (mode $m=1$). Drop keeps on rolling but eventually stop. The consecutive snapshots are extracted from movie SM1. (f-h)~The horizontal median cut views of the most unstable mode from the numerical stability analysis for some selected radii. The radii for the inner flow symmetry transitions from experiments and from stability analyses are plotted is (a) as black and orange lines, respectively.}
\end{figure}

{Leidenfrost drops levitate above a thin layer of vapor, of good insulating properties since $k_a<<k$, where $k_a$ and $k$ are the air and water thermal conductivities, provided in Table 1 of the SI.} Evaporation, which mainly takes place at the drop base, is thus markedly reduced and we verify here that the drop is at quasi-static equilibrium. A water drop with initial radius $R_0$ initially close to $\approx 4$~mm is deposited on a plate brought to 300$^{\circ}$C. We use a slightly curved substrate to immobilize the highly mobile liquid. The drop is observed using a top-view high speed camera, from which we extract $R$, the drop equatorial radius as evaporation proceeds. Figure \ref{fig:rvst} shows that $R$ decreases linearly with time $t$ at a rate $\mathrm{d}R/\mathrm{d}t\approx - 22 \rm\;\mu m/s$, so that the drop survives about $\tau_0\approx3$~minutes. \\
Temporal variations in $R$ are best fitted by a linear law, plotted as the blue dotted line and with equation:
\begin{equation}
R(t)=R_0(1-t/\tau_0),
\label{eq:fitRT}
\end{equation}
where $R_0 = 3.7\pm 0.1$~mm and $\tau_0 = 176.5$~s.
This time is much larger than the characteristic time of the internal motion $ R/V$, since tracers inside the liquid and at the drop surface reveal flow velocities $V$ as high as a few cm/s. The separation of time-scales $\tau_0 \gg R/V \approx 0.02$~s suggests that the evaporation-driven dynamics can be decoupled from the inner dynamics. As a result, a Leidenfrost drop can be considered in quasi-static equilibrium at any time, a key assumption to discuss the stability of Leidenfrost inner flows. 
Moreover, as will be discussed in \S \ref{sec:MaSt}, the instability develops faster than $\tau_0$, which supports the quasi-static stability analysis of the Leidenfrost drop.
\subsection{Leidenfrost drop shapes}
\label{sec:dropshape}
 A consequence of the timescale separation is that a given volume of liquid adopts the static shape of a non-wetting drop 
 \citep{Roman01}. If we denote $C(z)$ as the local curvature at a given height $z$ and $C_0$ as the curvature at the apex, the balance of hydrostatic and Laplace pressures can be written $C(z) = C_0 + z/\ell_c^2$. We introduce the curvilinear abscissa $s$, the horizontal radius at a given height $r(z)$ and the angle $\beta$ tangent to the interface, the previous equation can be recast into :
\begin{equation}
\frac{\sin \beta}{r} + \frac{\mathrm{d} \beta}{\mathrm{d}s} = C_0+\frac{z}{\ell_c^2}.
\label{eq:shape}
\end{equation}
A numerical integration of eq.(2) for $\beta$ ranging from 0 to $\pi$ {and $\mathcal{C}_0\ell_c$ ranging from 0.5 to 10, provides the drop shape for radii ranging from $R=0.9$~mm to 4.3~mm}, plotted as full lines in Figure \ref{fig:rvst}(b). These theoretical shapes are found to match the shapes obtained for water drops kept in place by a needle (see dotted lines), except at the drop north pole, where the needle locally forms a meniscus. Increasing the drop size tends to saturate the puddle height $H$ at its maximum value $H_{\max} \approx 2 \ell_c$, that is roughly 5~mm for water at 100$\C$. Drops smaller than the capillary length $\ell_c$ are quasi-spherical while drops larger than $\ell_c$ get flattened owing to gravity. As a consequence, evaporation induces geometric changes, particularly on the drop aspect ratio $2R/H$.
The presence of strong internal flows could in principle deform the liquid interface. The Reynolds number associated to the inner flows with typical velocity $V\sim$ cm/s and kinematic viscosity $\nu$ writes as $Re= R V/\nu$. For a millimetric drop, we have $Re \approx 10^3$, so that inertia overcomes viscosity.
The Weber number $We$, which compares inertia to the resisting capillarity is expressed as $We=\rho_0 V^2R/\sigma_0\sim10^{-2}$. Capillary therefore outbalances inertia, which justifies that the drop shape does not deviate from the static ones.
\subsection{Drop internal flow structure}
The apparent quietness of Leidenfrost drops does not reflect what really happens inside the liquid.
Side-viewed PIV measurements performed in a median plane of a water drop {containing} tracer particles have revealed strong internal flows, with velocities {of} a few cm/s. As the drop shrinks owing to evaporation, Leidenfrost flows undergo a series of successive symmetry breakings. This is further evidenced by focusing on the drop top surface. A Leidenfrost drop, kept on a concave substrate, is seeded with surface particles that have a greater affinity for the air interface. These hollow glass beads are i) pre-dispersed in water,ii) skimmed from the interface (where particles accumulate owing to a wetting or shape defect), and iii) introduced in a pre-dispensed Leidenfrost drop. Despite the apparent axisymmetry of the experiment, interfacial flow structures emerge, as illustrated by the top-views in Figures \ref{fig:rvst}(c-e) {and visible in the movie SM1).} When the drop has a radius $R>2.5$~mm, it hosts multiple vortices, which can be described by the azimuthal wavenumber $m \geq 3$ in a cylindrical coordinate system, by denoting ($r$, $z$, $\theta$) and using a periodic wavenumber expansion in $\theta$ as $e^{i m \theta}$, $m\in \mathbb{N}^*$.
Within the range $R=[1.8;2.5]$~mm, a mode $m=2$ clearly appears, while for smaller radius $R<1.5$~mm, the droplet rolls in an asymmetric fashion, corresponding to a mode $m=1$.
 Inner flows thus successively self-organize into 6 counter-rotating cells ($m=3$, Figure \ref{fig:rvst}(c)); 4 counter-rotating cells ($m=2$, Figure \ref{fig:rvst}(d)), and eventually a unique rolling cell ($m=1$, Figure \ref{fig:rvst}(e)). 
{We can notice that at some instance of movie SM1, the convective cells loose their organization and coherence. The flow structures seems to be transiently perturbed by the drop oscillation in the slightly curved well. However, the dominant modes reappear within a second as a hint of the robustness of the unstable modes.} The transition from $m=2$ to $m=1$ can be also visualized by side views, using PIV techniques, as in \cite{Bouillant2018a} (images reproduced in Figure \ref{fig:baseMa2410_exp}(a) and \ref{fig:EVPMa0}(a)) or even indirectly measured, as the drop acceleration $a$ (extracted from top-views, for water drops initially at rest) suddenly increases with the mode switching onto $m=1$.
 The experiment proposed by \citep{Bouillant2018a} is reproduced in the \citeyear{SI}, for plate temperatures ranging from $250$ to $450\C$. The accelerations $a$ of about 80 drops as a function of their radii $R$ exhibit similar jumps from $\sim$1~mm/s$^2$ (in the regime where drops are flattened by gravity), up to 60~mm/s$^2$ as $R$ decreases to $\sim 1$~mm. This indeed corresponds to entering the self-rotation and propelling regime \citep{Bouillant2018b}. 
The onset for the transition from $m=2$ to $m=1$, referred as $R_{2\rightarrow 1}$, is found to weakly depend on the plate temperature and to be within the interval $[1; 1.5]$~mm (see \S S6 of the \citeyear{SI}).
\subsection{Temperature gradient at the drop interface}
\label{subsec:temperaturedrop}
To test the aforementioned thermally-based instability scenario, we need to specify the thermal boundary conditions at the drop surface. The drop base is expected to be maintained at roughly the water boiling point ($T_b = 100\C$), while its apex is cooled down by the ambient air.
\begin{figure}[!htb]
\centering
\includegraphics[width=0.9\textwidth]{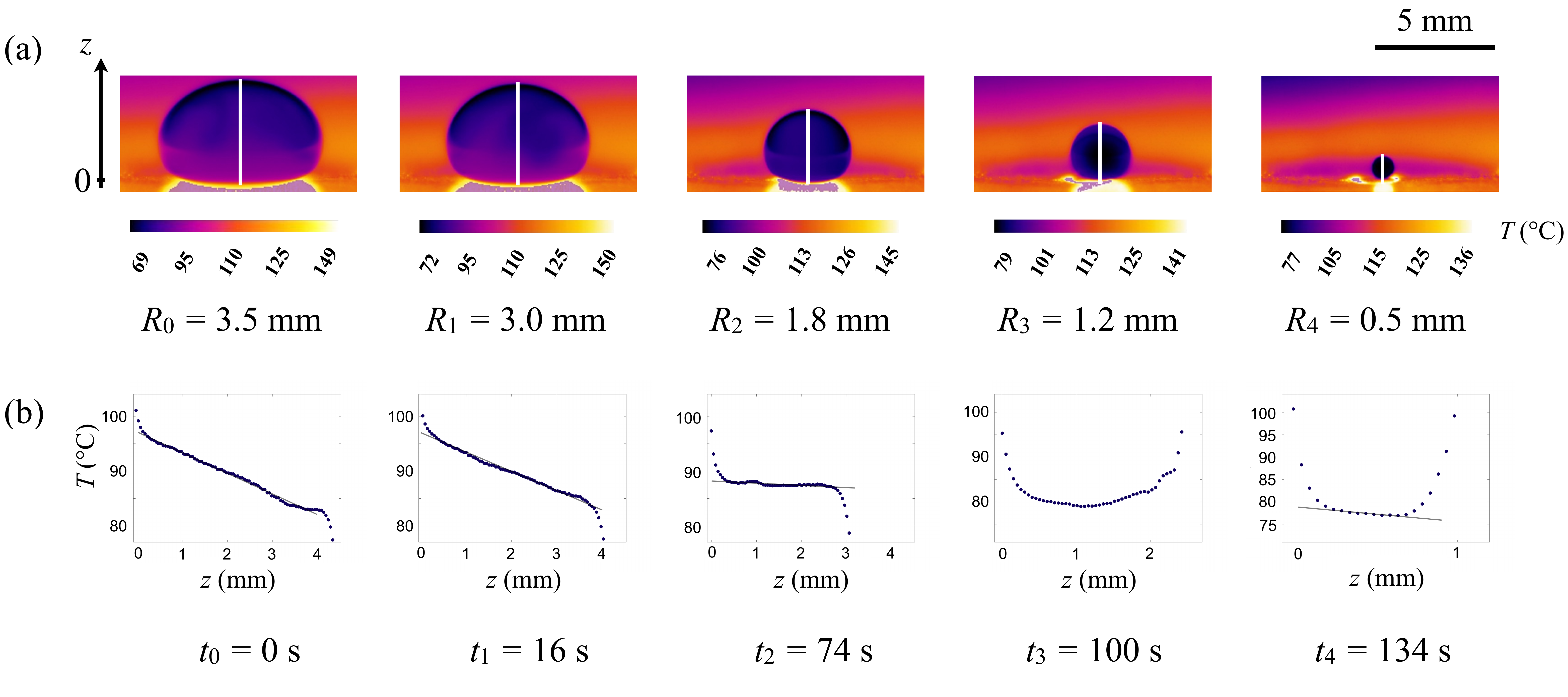}
\caption{\label{fig:expp}(a) Infrared side views of an evaporating Leidenfrost water drop deposited on a slightly curved surface of brass heated at 350$^{\circ}$C. Images, taken with a thermal camera give access to the surface temperature (calibration range from $-40^{\circ}$C to $+150^{\circ}$C to focus on water surface). Images are extracted from movie SM2. (b) Surface temperature $T$ of a given water drop along its central vertical axis $z$ (white dotted line), showing the change of temperature profile as the drop radius $R$ decreases.}
\end{figure}
The temperature field in a Leidenfrost drop is measured using an infrared camera (FLIR A600 series), calibrated on the temperature range $[-40 ;+ 150]\C$, only suitable to see water, and not the brass substrate. Water being opaque to infra-red wavelengths, this measurement provides the "skin" temperature $T$. Figure \ref{fig:expp} shows at $t = 0$, that is when $R = 3.5$~mm, $T$ linearly decreases with height $z$ (in millimeters) as $T = -3.75z + 97.0\C$, reaching a maximum $T_{\max} = 97.0\C$ at the drop base ($z = 0$). A temperature difference $\Delta T\approx 25\C$ thus develops along the drop. At $t = 16$~s, the gradient is slightly smaller with $T = -3.53z + 97.0\C$, and it keeps decreasing as $R$ decreases. These observations are confirmed by introducing a thermocouple inside the liquid (see Figure S1 in the (\citeyear{SI})).
For $t = 74$~s, when $R = 1.8$~mm, $T$ suddenly becomes homogeneous with $T \approx 88^\circ$C. As best visible in the supplementary movie SM1, this coincides with the moment where the flow switches to a symmetry $m=1$. At this transition, the drop starts to vibrate, which enhances mixing. For $t > 100$~s, $T$ thus becomes roughly homogeneous, with a minimum at the drop center (along the rolling axis), and a maximum at its periphery since fluid are periodically brought close to the hot plate.
In this rolling state, the liquid temperature is $T \approx 80^\circ$C, a value smaller than the boiling point of water, as consequences of i) the intensifying evaporation-driven cooling; ii) a reduction of the flattened area at the drop base from which water is heated, which scales as $R^4/l_c^2$ \citep{Mahadevan99}; iii) the temperature homogenization due to the {rolling-enhanced mixing}.
We now try to link the existence of such temperature distributions to the origin and structure of the internal flows.

\section{Theory}
\label{sec:Theory}
\subsection{Problem formulation}
\label{sec:Problem}
\begin{figure}[htb]
\centering
\includegraphics[width=0.6\textwidth]{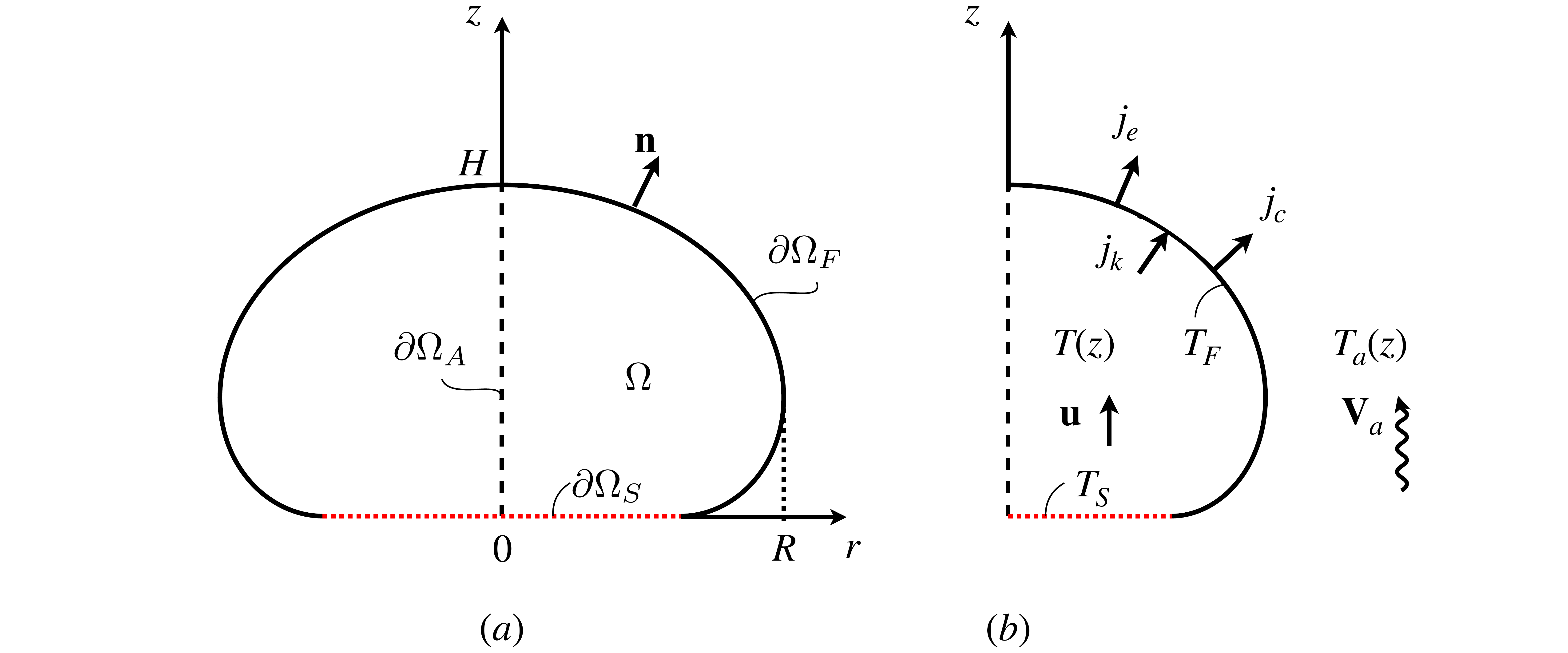}
\caption{\label{fig:Sketch_drop} Sketch of the problem. (a) The drops (domain $\Omega$) presents boundaries $\partial\Omega$ including the upper free surface $\partial\Omega_F$, the bottom interface $\partial\Omega_S$ and the axis of symmetry $\partial\Omega_A$.~(b) Thermal conditions.}
\end{figure}

Based on the experimental observations, we develop a minimal model assuming that i) the liquid adopts at any time the static shape of a non-wetting drop; ii) the baseflow inside a drop is steady; iii) the temperature at the bottom is fixed to the liquid boiling temperature; iv) the interaction between the liquid and the surrounding gas is not solved completely but modelled by heat transfer correlation laws applied on the side boundary.
We sketch in Figure \ref{fig:Sketch_drop}(a) the problem, where we represent from the side a static drop provided by eq.(\ref{eq:shape}). We denote by $\Omega$ the liquid domain, and by $\partial\Omega$ the boundaries, which are decomposed into $\partial\Omega_F$, the upper free interface and $\partial\Omega_S$, the bottom interface of the drop. We also introduce $\partial\Omega_A$ the vertical centerline of the drop, the axis of symmetry. Hence, the numerical computation is done only on the half domain $r=[0; R]$ as illustrated in Figure \ref{fig:Sketch_drop}(b).
The bottom interface $\partial\Omega_S$ is assumed to be isothermal at the temperature $T_s=100\C$ (the phase change of a pure body occurs at a given constant temperature), while the temperature on the side of the drop $\partial\Omega_F$ needs to be evaluated using the heat transfer balance. \\
\subsection{Governing equations}
\label{sec:GoveningEqn}
Under the Boussinesq approximation, the governing Navier-Stokes equation for the velocity fields $\BF{u}=[u_r,u_\theta,u_z]^\mr{T}$ and the temperature $T$ in the cylindrical coordinate $(r,\theta,z)$ defined in the domain $\Omega$ of boundaries $\partial\Omega$ reads, 
\begin{subequations}
\begin{alignat}{3}
\frac{\partial \BF{u}}{\partial t}+\bu \cdot \nabla \BF{u}&= -\frac{1}{\rho_0} \nabla p - \frac{\rho(T)}{\rho_0} \BF{g} + \nu \nabla^2 \bu \quad && \mr{in} \ \Omega,\\
\nabla \cdot \BF{u}&=0 \quad && \mr{in} \ \Omega, \\
 \frac{\partial T}{\partial t}+\BF{u} \cdot \nabla T &= \kappa \nabla^{2} T \quad && \mr{in} \ \Omega, 
\end{alignat}
\label{eq:dimGE}
\end{subequations}
where $\BF{g}$ is the gravitational acceleration in $z$ direction, $\nu$ the liquid kinematic viscosity and $\kappa$ its thermal diffusivity. 
The boundary conditions for $\partial\Omega_F$, $\partial\Omega_S$ and $\partial\Omega_A$ are respectively:
\begin{subequations}
\begin{alignat}{3}
\BF{S} \cdot \BF{n}&=\BF{S}_{\mathrm{gas}} \cdot \BF{n}-\sigma(\nabla \cdot \BF{n}) \BF{n}+(\BF{I}-\BF{nn}) \cdot \nabla \sigma \quad && \mr{on} \ \partial\Omega_F, \label{eq:dimBC1}
 \\
\BF{u\cdot n} &=0 \quad && \mr{on} \ \partial\Omega_F, \\
k \BF{n} \cdot \nabla T &= -h_{c} (T-T_{a}) - n' \mathcal{L} \quad && \mr{on} \ \partial\Omega_F, \label{eq:bcbasedimL2}\\
T &= T_0 \ && \mr{on} \ \partial\Omega_S, \label{eq:bcbasedim} 
\end{alignat}
\label{eq:dimBC2}
\end{subequations}
{with suitable boundary conditions on the flow axis of symmetry, $\partial\Omega_A$, detailed in \S \ref{sec:stability} for linear perturbations and reading  ${u}_r={u}_\theta = 0$ for the base axisymmetric case.}
$\BF{S}$ and $\BF{S}_{\mathrm{gas}}$ are the stress tensors in the liquid and the gas, respectively, $\sigma$ the surface tension, $\BF{n}$ the normal vector, $k$ the conductivity, $h_c$ the convective heat transfer coefficient in air, $n'$ the evaporation rate and $\mathcal{L}$ the latent heat. Equation \ref{eq:bcbasedimL2} expresses the heat flux {balance} at the interface stemming from the liquid ($j_k$) transferred into air ($j_c$) and into evaporative cooling ($j_e$), as schematized in Figure \ref{fig:Sketch_drop}(b). \\
Both the liquid density $\rho$ and the surface tension $\sigma$ are assumed to vary linearly with the temperature:
$\rho(T) = \rho_0 (1-\beta_w(T-T_0)), \ \sigma = \sigma_0 - \sigma_1 (T-T_0) ,$ where $\beta_w$ is the water thermal expansion coefficient, $\beta_w= {\rho_0}^{-1} {\partial \rho}/{\partial T}$ and $\sigma_1$ the surface tension variation with the temperature, $\sigma_1= -{\partial \sigma}/{\partial T}$.
We decompose now the temperature $T$ as:
\begin{equation}
T = T_0 - \Theta(r,z),
\end{equation}
where $T_0$ is the water boiling temperature taken as reference ($T_0=100\C$) at $z=0$ and $\Theta$ the deviation from $T_0$. The reference density $\rho_0$ and the surface tension $\sigma_0$ are defined at $T_0$.
With these definitions, the density and the surface tension write: 
\begin{align}
\rho(T) = \rho_0 \left(1-\beta_w \Theta \right) , \qquad
\sigma = \sigma_0 - \sigma_1 \Theta. \label{eq:surftension}
\end{align}
Denoting $p_0$ the hydrostatic pressure, the pressure $p$ inside the liquid decomposes as $p = p_0(z) + p_1(r,z)$.
The time, velocity and length are scaled with $H^2/\kappa, \kappa/H$ and $H$, respectively where $H$ is the drop height. The temperature is scaled with the temperature difference $\Delta T(=|\Theta_{z=0}-\Theta_{z=H}|)$ and the pressure is scaled with $\rho_0 \nu \kappa/ H^2$,
{which naturally appears when plugging eq.~\ref{eq:dimlessnum} in eq.~\ref{eq:dimGE}a, and comparing the viscous term to the pressure term,}
\begin{equation}
r= H\hat{r},\quad z= H \hat{z},\quad t = \frac{H^2}{\kappa} \hat{t}, \quad \bu = \frac{\kappa}{H} \hbu, \quad {\Theta} = \Delta T \hat{\Theta}, \quad p = \frac{\rho_0 \nu \kappa}{H^2} \hat{p}, \quad \nabla = \frac{1}{H}\hat{\nabla}.
\label{eq:dimlessnum}
\end{equation}
We provide in the \citeyear{SI} a Table S1 with the physical parameters relative to water, vapor, air and their interface relevant to describe our problem. We introduce the dimensionless numbers, Prandtl, Rayleigh, Marangoni, Biot and Sherwood numbers as:
\begin{equation}
{Pr}=\frac{\nu }{\kappa}, \quad {Ra}=\frac{\beta_w g \Delta T H^{3}}{\nu \kappa}, \quad Ma = \frac{\sigma_1 \Delta T H}{\rho_0 \nu \kappa}, \quad Bi = \frac{h_{c} H}{k}, \quad Sh = \frac{h_{m} H}{D_{va}}.
\label{eq:nodimnum}
\end{equation}
The governing equation (\ref{eq:dimGE}) can be now recast into
\begin{subequations}
\begin{alignat}{3}
Pr^{-1} \left( \frac{\partial \hbu}{\partial \hat{t}}+\hbu \cdot \hat{\nabla} \hbu \right)&= -\hat{\nabla} \hat{p}_1 + Ra \hat{\Theta} \BF{e}_z+ \hat{\nabla}^2 \hbu \quad && \mr{in} \ \Omega, \\
\hat{\nabla} \cdot \hbu&=0 \quad && \mr{in} \ \Omega,\\
\frac{\partial \hat{\Theta}}{\partial t}+ \hbu \cdot \hat{\nabla}\hat{\Theta} &= \hat{\nabla}^{2}\hat{\Theta} \quad && \mr{in} \ \Omega,
\end{alignat}
\label{eq:tth}
\end{subequations}
with the boundary conditions:
\begin{subequations}
\begin{alignat}{3}
-\hat{p}_1 \BF{n}+(\hat{\nabla} \hbu + (\hat{\nabla} \hbu)^T)\cdot \BF{n} &= - Ma ( \BF{I}- \BF{nn})\cdot \hat{\nabla} \hat{\Theta} \ && \mr{on} \ \partial\Omega_F, \\
\hbu \cdot \BF{n} &= 0 \ && \mr{on} \ \partial\Omega_F, \partial\Omega_S, \label{eq:bcun}\\
\BF{n} \cdot \hat{\nabla} \hat{\Theta} &= -Bi \left( \hat{\Theta}+\frac{T_0 -{T}_a}{\Delta T} \right) - \frac{n'\mathcal{L}H}{k}
\ && \mr{on} \ \partial\Omega_F, \\
\label{eq:bish}
\hat{\Theta} &= 0 \ && \mr{on} \ \partial\Omega_S.
\end{alignat}
\label{eq:bcbase}
\end{subequations}
and suitable symmetry conditions on $\partial \Omega_A$ (see 3.3 for details). The evaporation rate $n'$ is linked to the $Sh$ number, which represents the evaporative heat transfer coefficient. 
The heat exchanges on $\partial \Omega_F$, which determine $Bi$ and $Sh$ are modelled using the Ranz-Marshall correlation \citep{Ranz52,Bergman11} as detailed in the SI (\S S2), where the input ambient air temperature $T_a$ is also measured and provided in SI (\citeyear{SI}) (\S S4). 
\subsection{Stability analysis}
\label{sec:stability}
The steady toroidal baseflow ($\bu_b, p_b,\Theta_b$) is obtained by solving the nonlinear steady state solution of (\ref{eq:dimGE}) satisfying the boundary condition (\ref{eq:dimBC2}) using the Newton method. The baseflow is solved using the dimensional equations since some dimensionless numbers depend on $\Delta T$, which is also an unknown of the thermal problem.\footnote{Note that the non-dimensional equation can be resolved by defining $Ra$ and $Ma$ with the known parameters, i.e. $T_s$. However, we kept the classical definition of $Ra$ and $Ma$ which are with $\Delta T$.} 
We first compute the baseflows, estimate $\Delta T$ and deduce the dimensionless numbers. This approach differs from {thermoconvective studies on a flat plate or in a cylinder}. In these situations, {there is no radial pressure gradient} and the vertical gradient is solely balanced with the density, without inducing any velocity field. The temperature difference $\Delta T$ and {thermal dimensionless parameters are} control parameters. {In contrast, when the radial pressure gradient is nonzero (as in the Leidenfrost configuration, owing to the boundary conditions), it induces a steady flow with nonzero velocity}.
{Both $\Delta T$ and the dimensionless parameters are solutions of the problem.}
Assuming infinitesimal perturbations on the baseflow with the complex frequency $\omega$ and the azimuthal wavenumber $m$, the dimensionless flow field, pressure and temperature are decomposed using the normal mode expansion:
\begin{equation}
[\hat{u}_r,\hat{u}_\theta,\hat{u}_z,\hat{p}_1,\hat{\Theta}](r,\theta,z,t) = [{u}_r,{u}_\theta,{u}_z,{p},{\Theta}](r,z) \exp(\mr{i}m\theta - \mr{i} \omega t ) + c.c.,
\end{equation}
where $c.c.$ indicates complex conjugate. The linearized equation (\ref{eq:tth}) becomes
\begin{subequations}
\begin{alignat}{3}
Pr^{-1} \left( \mr{i} \omega \bu +\bu_b \cdot \nabla_m +\bu \cdot \nabla_0 \bu_b \right)&= -\nabla_m {p} + Ra \Theta \BF{e}_z + \nabla^2_m \bu \quad && \mr{in} \ \Omega, \\
{\nabla}_m \cdot \bu&=0 \quad && \mr{in} \ \Omega, \\
\mr{i} \omega \Theta + \bu_b \cdot {\nabla_m} \Theta + \bu \cdot {\nabla_0} \Theta_b &= {\nabla}_m^{2}\Theta\quad && \mr{in} \ \Omega,
\end{alignat}
\label{eq:tthLNS}
\end{subequations}
where $\nabla_m$ represents the derivative in $\theta$ is replaced by $\mr{i}m$. The boundary conditions on $\partial\Omega_F$ are
\begin{subequations}
\begin{alignat}{3}
-{p} \BF{n}+({\nabla}_m \bu + ({\nabla}_m \bu)^T)\cdot \BF{n} &= - Ma ( \BF{I}- \BF{nn})\cdot {\nabla}_m {\Theta} \quad && \mr{on} \ \partial\Omega_F, \\
\bu \cdot \BF{n} &= 0\quad && \mr{on} \ \partial\Omega_F, \\
\BF{n} \cdot {\nabla}_m {\Theta} &= -Bi {\Theta}\quad && \mr{on} \ \partial\Omega_F.
\end{alignat} 
\end{subequations}
Note that the perturbation temperature is now only affected by the $Bi$ as the other heat transfer coefficients are constant and contribute only in the zero order baseflow equation. One could also linearize the last term in (\ref{eq:bcbasedimL2}) and include in the stability analysis, but the effect on the stability results is negligible \citep{Yim21}.
The condition prescribed {on the drop axis of symmetry $\partial\Omega_A$ (illustrated in figure \ref{fig:Sketch_drop})} depends on $m$, in particular on the mode symmetry. The following conditions are thus used as in \cite{Batchelor62}:
 \begin{subequations}
\begin{alignat}{3}
m=0,& \quad u_r=u_\theta = \frac{\partial u_z}{\partial r}= \frac{\partial \Theta}{\partial r}=0 \quad && \mr{on} \ \partial\Omega_A, \\
m=1,& \quad u_z=\Theta = p = \frac{\partial u_r}{\partial r}=0=\frac{\partial u_\theta}{\partial r}=0 \quad && \mr{on} \ \partial\Omega_A, \\
m\geq2,& \quad u_r=u_\theta = u_z = \Theta =p= 0 \quad && \mr{on} \ \partial\Omega_A. 
\end{alignat} 
\end{subequations}
Finally, we prescribe on $\partial\Omega_S$, a Dirichlet condition for the temperature and a free slip condition for the velocity since the bottom surface is not in contact with the plate: 
\begin{subequations}
\begin{alignat}{2}
\bu \cdot \BF{n} &= 0\quad && \mr{on} \ \partial\Omega_S, \\
\Theta &= 0 && \mr{on} \ \partial\Omega_S.
\end{alignat} 
\end{subequations}
\subsection{Numerical methods}
All the numerical analyses are performed using FreeFEM++ software \citep{Hecht12} for axisymmetric cylindrical coordinates $(r, z)$. The velocity, pressure and temperature are discretized with Taylor-Hood P2, P1 and P2 elements, respectively. 
The typical number of triangles is $\sim 10^4$. The linear equations and the eigenvalue problem are solved using UMFPACK library and ARPACK shift-invert method, respectively. Starting with the initial radius $R=3.5$~mm, the nonlinear solution of (\ref{eq:dimGE}) is solved for the given water properties. Once the solution for one radius is found, it is used as the initial guess for the smaller radius. Within 5 iterations, the L2 norm residual becomes smaller than $1 \cdot 10^{-8}$.
 The zero normal velocity condition on the free surface is applied using the Lagrange multiplier method \citep{Babuvska73,Yim21}. 

\section{Numerical results}
\label{sec:result}
\subsection{Pure buoyancy induced flow ($Ma=0$)}
\subsubsection{Baseflow ($Ma=0$)}
\label{sec:baseflowMa=0}

\begin{figure}[!h]
\centering
\includegraphics[width=0.95\textwidth]{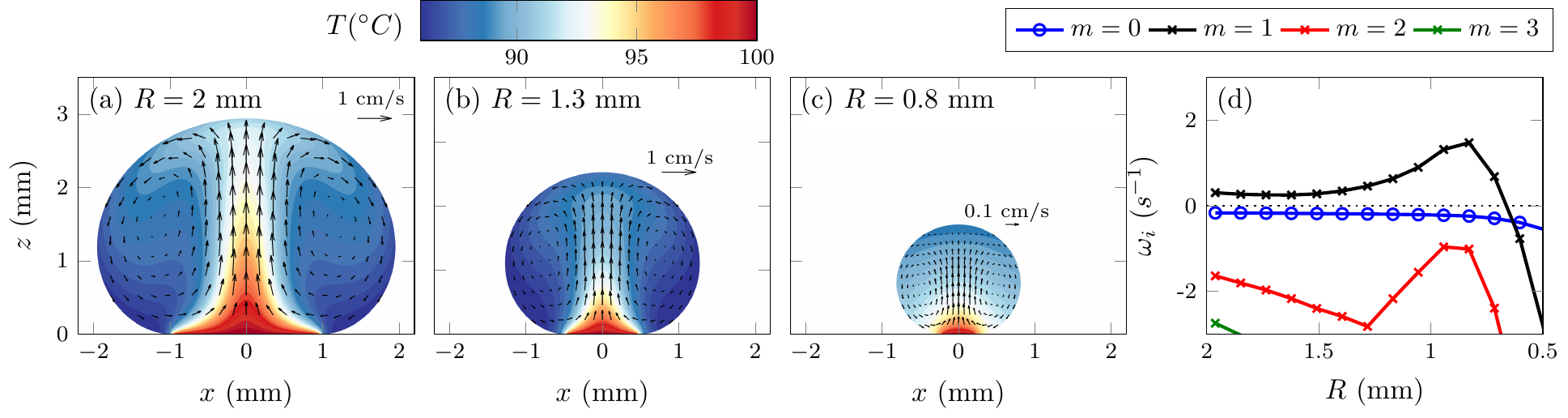}
\caption{\label{fig:baseMa0} Baseflows for $Ma=0$: (a) $R=2$~mm, (b) $R=1.3$~mm and (c) $R=0.8$~mm. The temperature and velocity fields are shown in color and with arrows. (d) The spectrum of growth rate $\omega_i$ with $R$.\\}
\end{figure}

Let us first consider the case of pure buoyant flows, neglecting thermocapillary effects. The baseflow is thus computed following (\ref{eq:dimGE}) while setting the superficial stress on $\partial\Omega_F$ to zero. We restrict this parametric study in $R$ to the limit $R \lesssim 2$~mm, since according to IR measurements (Figure~\ref{fig:expp}), the drop surface temperature tends to homogenize, suppressing Marangoni surfaces flows. Figure \ref{fig:baseMa0} shows the baseflow obtained in this limit of $Ma=0$, for drop radii $R=2$~mm, $R=1.3$~mm and $R=0.8$~mm, respectively. In the absence of surface tension gradient, the baseflow exhibits pure thermal convection: the flow rises along the center axis, warmer since it is insulated from the drop interface, and descends along the side of the {drop}, where the drop is cooled.
The inner velocities for purely buoyant flows ($\sim 1$~cm/s) underestimate the experimental observation ($\sim 5$~cm/s) of the similar size of drop (see Figure \ref{fig:EVPMa0}a,c for the experimental measurements). \\

\subsubsection{Stability analysis ($Ma=0$)}
\label{sec:StabilityAnalysisMa=0}
\begin{figure}[!h]
\centering
\includegraphics[width=1\textwidth]{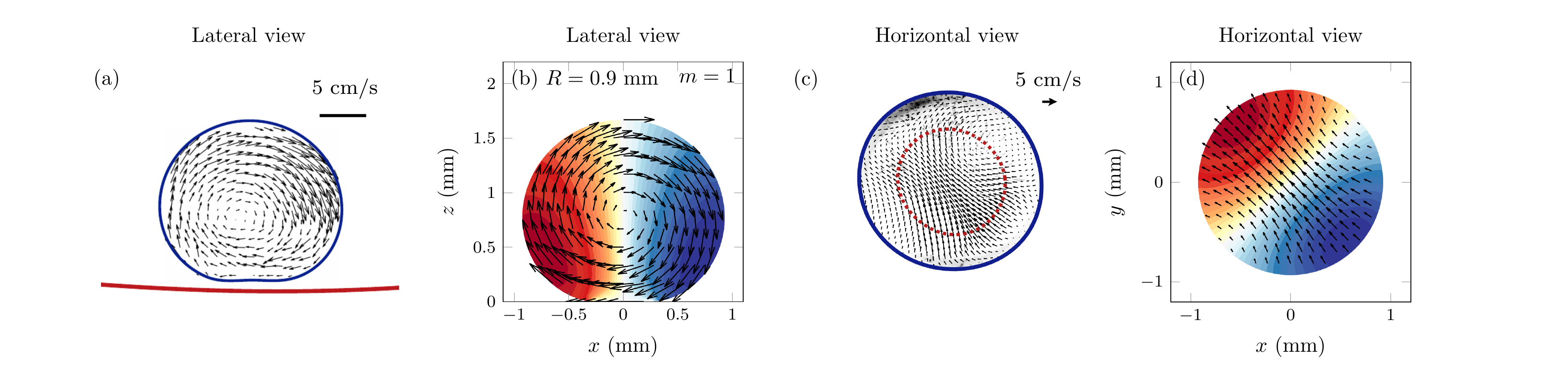}
\caption{\label{fig:EVPMa0} (a,c) {Velocity} fields within a droplet with $R \sim 0.9$~mm deduced from PIV measurements. (b, d) Corresponding flow fields deduced from the numerical stability analysis in the absence of Marangoni effects ($m=1$, $Ma=0$, $Ra=6.1 \cdot 10^3$). Velocity arrows are plotted within {the lateral (a,b) and horizontal (c,d) planes}. The red dashed line in (c) indicates the area of the flattened base of the drop on which the bottom camera focuses. Color in (b,d) indicates the perturbed temperature field (normalized with its maximum value).}
\end{figure}

The stability of purely thermobuoyant flows is herein considered. Figure~\ref{fig:baseMa0}(d) shows the growth rate $\omega_i$ (imaginary part of the complex frequency $\omega$) as a function of $R$ for the modes $m=0,1,2,3$. A positive growth rate indicates the grow of the perturbations leading to the instability. As shown in Figure~\ref{fig:baseMa0}(d), the $m=1$ mode is only unstable mode for $R>0.6$~mm. The frequency $\omega_r$ of this mode (not shown) is zero, corresponding to a steady unstable mode.
The corresponding Rayleigh number $Ra$ decreases from $20000$ to 10 as $R$ decreases from 2 to 0.5~mm (see Fig.~S8b in \citeyear{SI}), {reaching the value $Ra\sim 630$ when $R\sim 0.6$~mm, for which flows get stable}. 
Interestingly, this limit also corresponds to the radius where the droplet ability to self-rotate and propel disappears, as noticeable in the last stage of movie SM1. Both the measured propelling acceleration and the droplet base asymmetry vanish below $R \lesssim 0.6$~mm until it stops (the measurement in Figure \ref{fig:rvst}(a) then ceases). This suggests that thermobuoyant effects become stable to non-axisymmetric disturbances and flows stabilize as the {drop} size reduces below a critical value.
 Figure \ref{fig:EVPMa0} compares the $m=1$ unstable mode to the flow fields obtained in a Leidenfrost drop with $R \sim 0.9$~mm. The mode $m=1$ with a structure describes well the solid-like rolling motion in the experimental observation. \\
 
Although the $m=1$ unstable mode with $Ma=0$ represents well the rolling motion of small drops observed in the experiments, it fails to describe the presence of higher azimuthal modes for large drops. This prompts us to look at the $Ma\neq 0$ case for larger radii.

\subsection{Reduced Marangoni approximation}
\label{sec:ReducedMa}
A major challenge for computing the baseflow in a Leidenfrost drop is to be able to predict the surface tension distribution. When we take the exact surface tension temperature dependence $\sigma_1$ provided in the Table S1 (\citeyear{SI}), the Marangoni flows are largely overestimated compared to experimental observations (see Figure S3 of the \citeyear{SI}).
It is known however that Marangoni effects are very often markedly reduced \citep{Hu05,Hu06,Dhavaleswarapu10} or even almost absent \citep{Marin11,Gelderblom12,Dash14}.
 Based on measured quantities, we first try to correct the surface tension temperature dependence $\sigma_1$. To that end, we evaluate the Rayleigh $Ra$ and Marangoni $Ma$ numbers, which compare the buoyancy and surface tension stresses in comparison to inertia, respectively, as defined in (\ref{eq:nodimnum}). Using the parameters documented in Table S1 of the \citeyear{SI}, they culminate to $Ra \sim 1.8 \cdot 10^5$ and $Ma \sim 3.6 \cdot 10^5$ for $\Delta T = 25\C$, $H=4$~mm, $R=3.5$~mm. 
Both values greatly exceed expected critical values for the onset for Rayleigh-B\'enard and Marangoni instabilities (see Table S2 in the SI) -- typically $Ra_c \sim O(10^3)$ and $Ma \sim O(10^2)$, for similar geometries \citep{Chandrasekha1961}. 
 As detailed in \S S7 of the \citeyear{SI}, the effective surface tension variation $\sigma_{1,\rm eff}$ (and thus $Ma_{\rm \rm eff}$) is determined using numerical analysis. We select $\sigma_{1,\rm eff}$ that best describes the temperature difference $\Delta T$ within the liquid (Fig~\ref{fig:expp}), as well as the flow velocities, typically 5~cm/s. The temperature profiles reported in Figure \ref{fig:expp} (left panels, $R=3.5$~mm and $R=3.0$~mm) are best represented by the curve with $\sigma_{1,\rm eff}=4 \cdot 10^{-5} \sigma_1$ as shown in Figure S7 of the SI for the surface temperature and Figure \ref{fig:baseMa2410_exp} for velocity field.  {The surface stress seems to be reduced by a few order of magnitude.  This has also been reported in \cite{Savino2002,Hu02,Hu05Ma0,Hu06},  where it is ascribed to surface contamination,  or to the fact that at large Marangoni surface stress,  it gets moderated by dissipation or by transport-limited properties of the fluid, reducing the achievable velocities (see Fig.6 of the SI).}
 In the following stability analysis, we thus use this reduced surface tension variation by adjusting $\sigma_1$ to $\sigma_{1,\rm eff}=4 \cdot 10^{-5} \sigma_1$, which is the only tunable parameter of our study (all other physical parameters are kept exact values of given thermal properties, provided in the Table S1).
\subsection{Thermocapillary flow}
\label{sec:Thermocapillaryflow}
\subsubsection{Baseflow with effective Ma}
\label{sec:baseflowMarmeff}
\begin{figure}[!htb]
\centering
\includegraphics[width=0.9\textwidth]{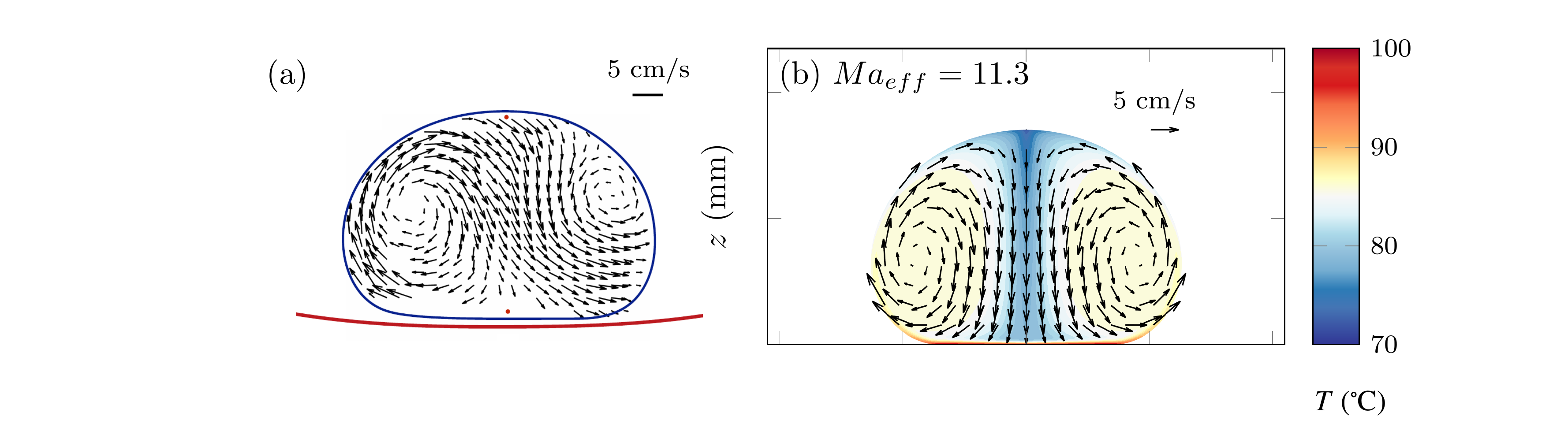}
\caption{\label{fig:baseMa2410_exp} (a) PIV measuments in a drop with $R=2.5$~mm \citep{Bouillant2018a}.~(b) Baseflow for $Ma_{\rm eff}=11.3$ ($\sigma_{1,\rm eff} = 4\cdot 10^{-5} \sigma_1$). The colormap and arrows give the inner temperature and velocity.}
\end{figure}

The baseflow in Leidenfrost drops is now computed as in \ref{sec:baseflowMa=0}, adding to thermobuoyant effects reduced thermocapillary effects ($\sigma_{1,\rm eff}=4\cdot 10^{-5} \sigma_{1}$ and $Ma_{\rm eff}$) and shown in Figure \ref{fig:baseMa2410_exp} compared to the experimental measurement (Figure \ref{fig:baseMa2410_exp}a). As the surface tension gradient induces Marangoni flow from low surface tension to the higher one along the surface boundary, the flow direction is opposed than purely buoyant flow: it sinks on the center-line and rises along the surface. The typical velocity magnitude is about $\sim 5$~m/s which is similar to the experimental observation,which corroborates the choice of effective thermocapillary gradient done by tuning the temperature distribution in the previous section \ref{sec:ReducedMa}.\\

\subsubsection{Stability analysis with effective Ma}
\label{sec:MaSt}
\begin{figure}[!ht]
\centering
\includegraphics[width=0.48\textwidth]{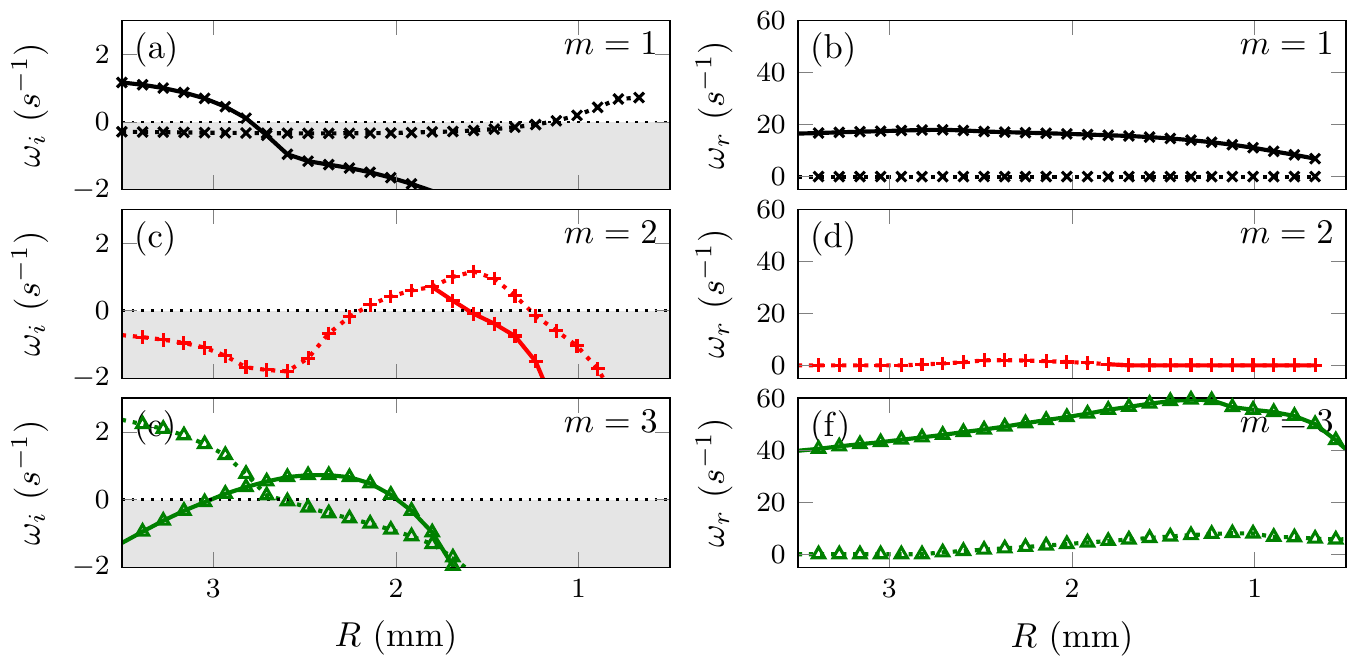}
\includegraphics[width=0.48\textwidth]{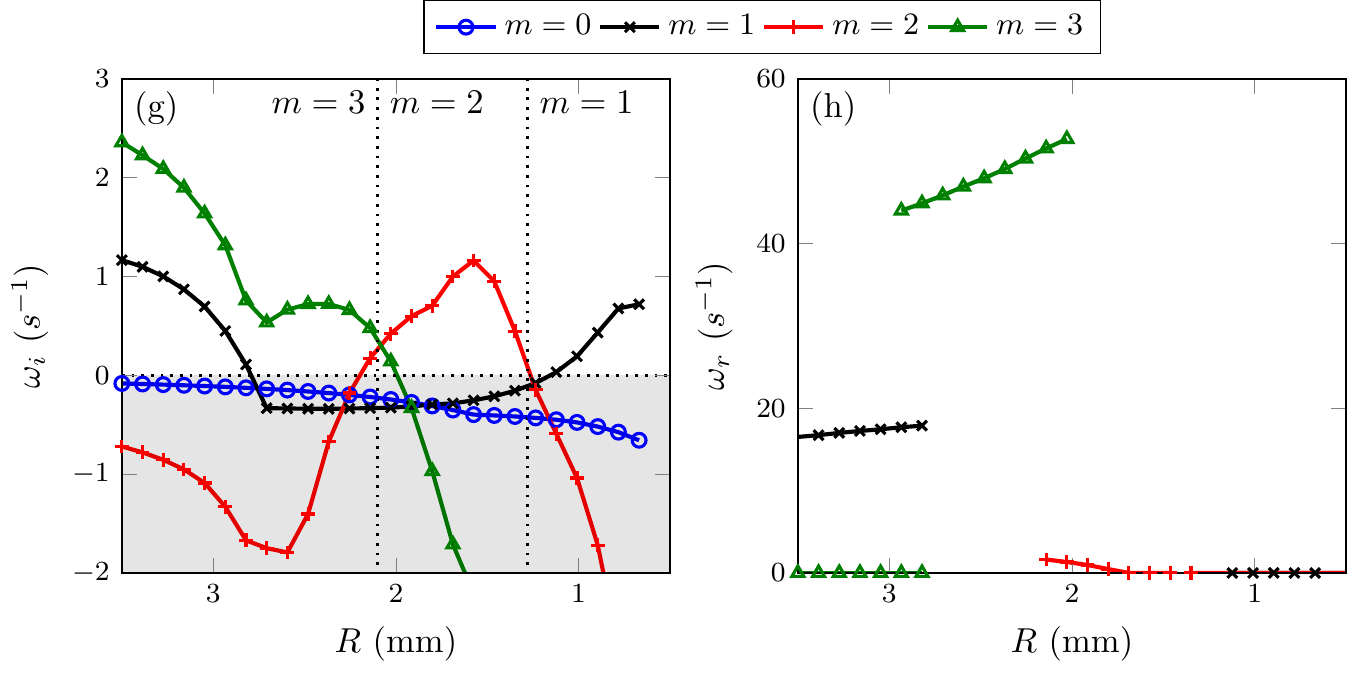}
\caption{\label{fig:spectMa0_all} Growth rates and the corresponding frequencies for (a,b) $m=1$, (c,d) $m=2$ and (e,f) $m=3$ for the two least unstable modes as a function of decreasing radius. The dominant modes for each $m$: (g) growth rate and (h) corresponding frequency.}
\end{figure}
Figure \ref{fig:spectMa0_all} shows the dominant eigenvalues as a function of decreasing radius (with $\sigma_{1,\rm eff}=4\cdot 10^{-5} \sigma_{1}$). Compared to the pure buoyant flow, there exist several unstable azimuthal modes ranging from $m=1$ to 3. Figures \ref{fig:spectMa0_all}(a,c,e) show the two least stable growth rates $\omega_i$ and their corresponding frequencies $\omega_r$ are shown in Figures \ref{fig:spectMa0_all}(b,d,f) for azimuthal wavenumbers $m=1,2$ and 3. For all modes $m=1,2,3$, there exist two branches of unstable modes: one with $\omega_r=0$ (steady) and other with $\omega_r \neq 0$ (unsteady). 
 The $m=1$ mode (Figure \ref{fig:spectMa0_all}a,b) is unstable both at large radius $R>2.8$~mm and small radius $R<1.3$~mm, but it shows a window of stability for intermediate values of the radius. The prevailing unstable mode for $R>2.8$~mm is unsteady, with frequency $\omega_r \sim 10\ \mr{s^{-1}}$, while for $R<1.3$~mm, it becomes steady. \\
 Mode $m=2$ (Figure \ref{fig:spectMa0_all}c,d) is only unstable in the intermediate radius range $R=[2.3 ; 1.3]$~mm. At radius $R\sim 2$~mm, the unsteady branch dominates, yet, with a small frequency, but the steady branch takes over at smaller $R$. 
 Mode $m=3$ (Figure \ref{fig:spectMa0_all}e,f) is unstable when $R<2.8$~mm. Its steady branch dominates at large $R$ but it becomes unsteady as the radius gets closer to $R\approx 2.8$~mm. The unsteady branch of mode $m=3$ reaches a maximum growth rate around $R=2.4$~mm. \\
 Finally, Figures \ref{fig:spectMa0_all}g,h collect and retain only the most unstable mode $m$ and their corresponding frequencies. We note in Figure \ref{fig:spectMa0_all}g that the axisymmetric mode $m=0$ is always stable ($\omega_i<0$). The dependence of the dominant azimuthal mode on radius thus becomes explicit: 
 for $R > 2.1$~mm, the mode $m=3$ is the most unstable (first steady and then oscillatory). As the radius decreases further, the $m=3$ mode becomes stable around $R=2$~mm. At $R=2.2$~mm, the mode $m=2$ starts to grow and becomes the only unstable mode in the range of radii $R=[2; 1.3]$~mm, with the maximum growth rate at $R=1.6$~mm. For $R<1.3$~mm, the steady mode $m=1$ takes over as the dominant unstable mode. 
This trend is very similar to the apparent mode transition in experiments, which is $m\geq 3$ for $R \gtrsim 2.8$~mm, $m=2$ for $R \approx [2.5 ; 1.8]$~mm and $m=1$ for $R \approx 1.5$~mm, as shown in Figure~\ref{fig:rvst}. \\

\begin{figure}[!htb]
\centering
\includegraphics[width=0.85\textwidth]{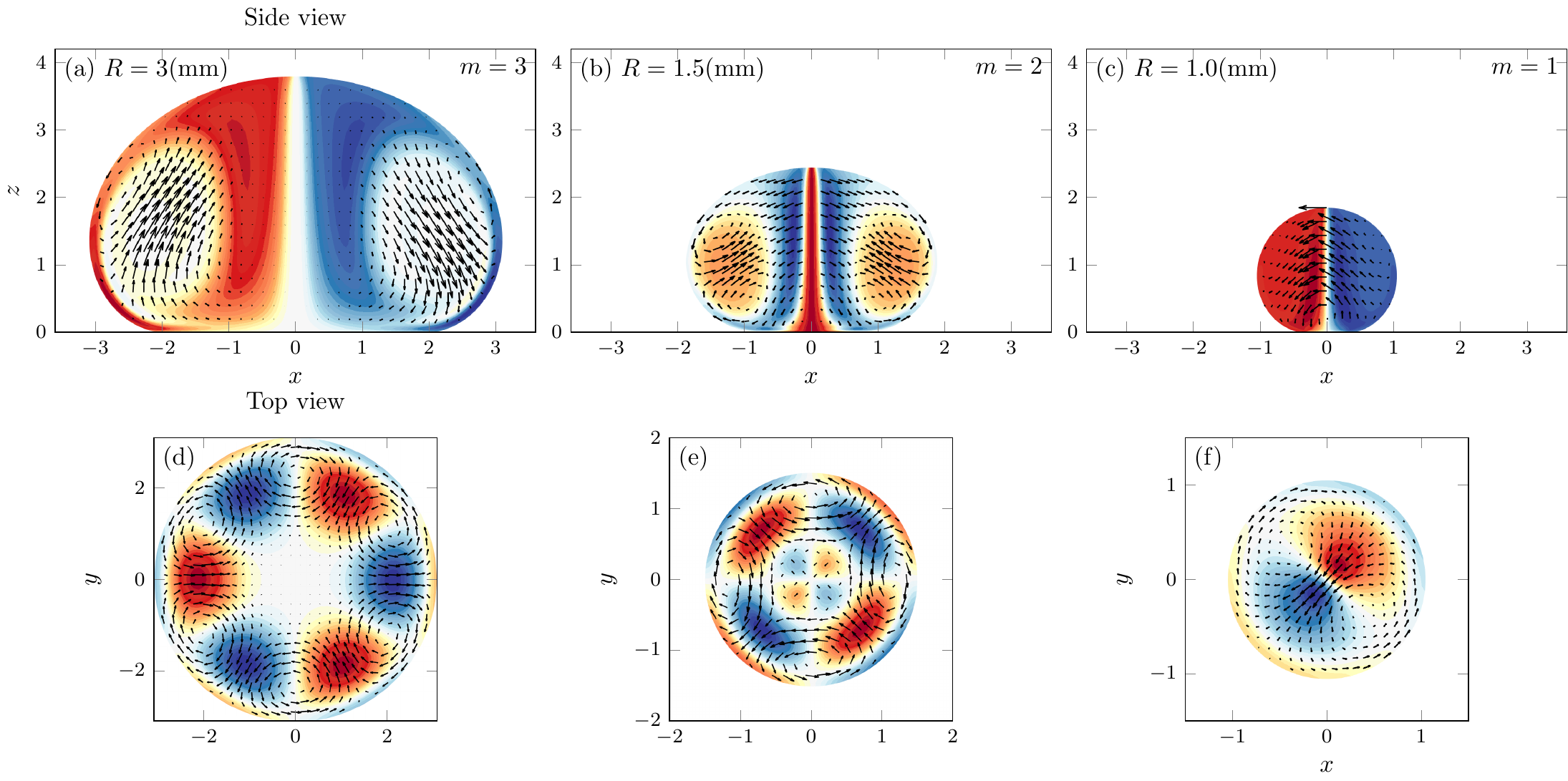}
\caption{\label{fig:modesMa4e-5} Eigenmodes of the most unstable modes at radii (a,d) $R=3$~mm ($Ma_{\rm eff}=12, Ra=1.5\cdot10^5$), (b,e) $R=1.5$~mm ($Ma_{\rm eff}=9.6, Ra=5\cdot10^4$) and (c,f) $R=1$~mm ($Ma_{\rm eff}=8, Ra=2.4\cdot10^4$). Colors indicate temperature perturbations and the arrows show the in-plane velocity perturbations normalized with their maximum real values. The top view is a plane cut at the maximum radius $R=R_{\max}$. }
\end{figure}
A growth rate $\omega_i \sim1~\mr{s^{-1}}$ implies that within 2.3 s, the perturbation amplitude becomes 10 times larger than its initial value. For all modes, we verify that the growth rate is much faster than the {drop} evaporation rate $\omega_i \gg (1/R)\mr{d}R/\mr{d}t$, underlying the quasi-steady assumption in \S \ref{sec:quasisteady}.\\
Figure \ref{fig:modesMa4e-5} illustrates the eigenvectors of the most unstable mode at radii $R=3,\ 1.5$ and $1$~mm. 
The top panel displays side-views while the right one shows top-views: $xy$ plane cut at the maximum radius $R=R_{\max}$.
The colors indicate temperature perturbations and the arrows are the in-plane velocity perturbations.
For $R=3$~mm (Figure \ref{fig:modesMa4e-5}ad), the mode $m=3$ dominates. Temperature perturbations are localized along the interface and the central axis of the drop while velocity perturbations are localized where the temperature perturbations are the weakest. The top view shows three counter-rotating vortex pairs, very similar to our observations (Figure \ref{fig:rvst}c). 
For $R=1.5$~mm (Figure \ref{fig:modesMa4e-5}be), the mode $m=2$ becomes the most unstable thus dominant mode. Temperature perturbations are maximum near the central axis, while velocity perturbations remain localized where temperature perturbations are weak. The top view shows 4 vortex cells (or 2 vortex pairs), in close agreement to the experiments (Figure \ref{fig:rvst}d).
For $R=1$~mm (Figure \ref{fig:modesMa4e-5}cf), the displacement mode $m=1$ is the most unstable. Temperature perturbations are null on the central axis, where velocity perturbations are maximum.
However, its structure does not match the solid-like rolling motion reported in \cite{Bouillant2018a} and illustrated in the top of Figure \ref{fig:rvst}e. We interpret this at a consequence of the temperature homogenization evidenced at small $R$ in Figure~\ref{fig:expp}. The temperature difference at the liquid surface indeed tends to vanish as $R$ becomes smaller than 1.8~mm. We thus expect thermocapillary effects to weaken and eventually vanish. We provide in Figure S8 (\citeyear{SI}), the (reduced) Marangoni and Rayleigh numbers with $R$, showing that both effects weaken as the drop shrinks in size, with a more abrupt decrease of thermocapillary effects than thermobuoyant effects, prompting us to use our predictions for the case $Ma=0$ when $R$ becomes sufficiently small.
The observed rolling motion is thus better captured by the eigenmode for $Ma=0$ as shown in Figure \ref{fig:EVPMa0} than the one with $Ma \sim O(10)$ in Figure \ref{fig:modesMa4e-5}e. 
Moreover, the drop shape is, as yet, steady, but the internal flows could be coupled to the drop envelop deformation. This extension of our model would enable the exploration of the limit of even larger drops ($R \gtrsim 5$~mm), where star-pulsations appear. Finally, it would be interesting to see how our model applies to the inverse-Leidenfrost situation \citep{Gauthier2019}, for which a rolling mode $m=1$ seems dominant in the millimetric, quasi-spherical droplet. Their configuration is however essentially different since temperature gradients are reversed, the drop shape remains quasi-spherical on the deformable bath, and, as soon as the drop freezes,  both the thermo-buoyant and thermo-capillary flows should extinguish.

%
%
%
%

\bibliographystyle{jfm}
\bibliography{biblio_all}

\end{document}